\documentclass[12pt,preprint]{aastex}

\newcommand{\etal}{{et al.}}
\newcommand{\eg}{{\it e.g.,}}
\newcommand{\ie}{{\it i.e.,}}

\begin{document}

\title{Two Active Nuclei in 3C\,294\footnotemark[1]}

\footnotetext[1]{Based in part on observations made with the NASA/ESA Hubble
Space Telescope, obtained at the Space Telescope
Science Institute, which is operated by the Association of Universities
for Research in Astronomy, Inc., under NASA contract NAS 5-26555. These
observations are associated with proposal no. 08315. Based in part on data collected 
at Subaru Telescope, which is operated by the National Astronomical Observatory of Japan. 
Some of the data presented herein were obtained at the W.M. Keck Observatory, which is 
operated as a scientific partnership among the California Institute of Technology, the 
University of California and the National Aeronautics and Space Administration. The 
Observatory was made possible by the generous financial support of the W.M. Keck 
Foundation. Some of the data were also obtained from the Chandra Data Archive, part of the 
Chandra X-Ray Observatory Science Center, which is operated for NASA by the Smithsonian 
Astrophysical Observatory}

\author{Alan Stockton}
\affil{Institute for Astronomy, University of Hawaii, 2680 Woodlawn
 Drive, Honolulu, HI 96822}

\author{Gabriela Canalizo\altaffilmark{2}}
\affil{Institute of Geophysics and Planetary Physics, Lawrence Livermore
National Laboratory, 7000 East Avenue, L413, Livermore, CA 94550}

\author{E. P. Nelan}
\affil{Space Telescope Science Institute, 3700 San Martin Drive, Johns
Hopkins University Homewood Campus, Baltimore, MD 21218}

\author{Susan E. Ridgway}
\affil{Department of Physics and Astronomy, Johns Hopkins University,
Homewood Campus, Baltimore, MD 21218}

\altaffiltext{2}{Current address: Institute of Geophysics and Planetary Physics and
Department of Earth Sciences, University of California, Riverside, CA 95521}

\begin{abstract}
The $z=1.786$ radio galaxy 3C\,294 lies $<10$\arcsec\ from a 12th mag star and has been the target of at least three previous investigations using adaptive-optics imaging. A major problem in interpreting these results is the uncertainty in the precise alignment of the radio structure with the $H$ or $K$-band AO imaging. Here we report observations of the position of the AO guide star with the Hubble Space Telescope Fine Guidance Sensor, which, together with positions from the U. S. Naval Observatory's UCAC2 catalog, allow us to register the infrared and radio frames to an accuracy of better than 0\farcs1. The result is that the nuclear compact radio source is not coincident with the brightest discrete object in the AO image, an essentially unresolved source on the eastern side of the light distribution, as \citet{qui01} had suggested. Instead, the radio source is centered about 0\farcs9 to the west of this object, on one of the two apparently real peaks in a region of diffuse emission. Nevertheless, the conclusion of \citet{qui01} that 3C\,294 involves an ongoing merger appears to be correct: analysis of a recent deep Chandra image of 3C\,294 obtained from the archive shows that the nucleus comprises two X-ray sources, which are coincident with the radio nucleus and the eastern stellar object. The X-ray/optical flux ratio of the latter makes it extremely unlikely that it is a foreground Galactic star.
\end{abstract}


\section{Introduction}

There are substantial reasons to believe that radio galaxies at high redshifts comprise at least a significant part of the parent population of massive central cluster galaxies at the current epoch. Given that the growth of bulges and the growth of supermassive black holes are likely to be intimately related \citep{kor95,mag98,geb00,fer00}, the most powerful active nuclei will almost certainly be found in massive galaxies. At high redshifts, such galaxies are expected to have formed almost exclusively in regions of high overdensity, in which the processes of galaxy formation will have proceeded most rapidly, and which will also be likely seedbeds for protoclusters.

Recent observational evidence supports this picture (see \citealt{bes00} for a review of earlier results). In particular, although high-redshift radio sources are found in a variety of environments, evidence for clustering around many of them has been found in both optical/IR \citep[\eg][]{bes03} 
and X-ray \citep{pen02} surveys. In some cases, there are indications that the radio source is the central galaxy in the cluster, either from an overdensity of galaxies at small scales around it \citep[\eg][]{bes03}, or from the surface-brightness profile of the galaxy itself \citep{bes98}. This means that the study of the morphologies and other properties of high-redshift radio galaxies that appear to be the dominant members of clusters or protoclusters can give us insights into the formation of the most massive galaxies in the present-day universe.

One such case is the $z=1.786$ radio galaxy 3C\,294, which \citet{tof03} have found to be surrounded by an overdensity of faint red galaxies. This object has been a favorite target for adaptive
optics (AO) systems on large telescopes both because it is one of the most
powerful radio galaxies in the observable universe and because its optical center lies
$<10$\arcsec\ from a 12th mag star.  
The first AO imaging of this object, using the University of Hawaii curvature-sensing 
AO system {\it Hokupa`a} on the
Canada-France-Hawaii Telescope (CFHT), was reported by \citet{sto99}, who found a clumpy 
structure in the $K'$ band and suggested that the various clumps might be dusty subunits in the 
process of merging and illuminated by a hidden nucleus to the south of most of the observed
structure. They also emphasized the uncertainty in the position in the nucleus.

\citet{qui01} used the AO system on the Keck II telescope to observe 3C\,294 in the $H$ and
$K'$ bands.  They found an essentially stellar profile ($<50$ mas FWHM) for the eastern 
component, separated by $\sim1\arcsec$ from the diffuse western component. Taking the
USNO-A2.0 position for the AO guide star, they concluded that the
active nucleus is coincident with the stellar eastern component.  They interpret the structure
as an ongoing merger of two galaxies, with the active nucleus associated with the fainter
galaxy.

\citet{ste02} found a position for the radio source similar to that of \citet{sto99}, but they did not
state how they calculated it. Their AO imaging at $H$ and $K$, obtained with {\it PUEO}  on the
CFHT, showed 3 main components: their component $c$ corresponds to the stellar eastern
component of \citet{qui01}; their $a$ and $b$ are two subclumps in the diffuse western 
component.  The position they gave for the radio source did not correspond to any of the
features visible at $H$ or $K$.

Although these 
observations give the highest optical/IR resolutions yet achieved on any powerful 
high-redshift radio galaxy, the interpretation of the imaging remains ambiguous
because the precise location of the radio nucleus remains uncertain.
By far the largest part of this uncertainty is that of the position of the AO 
guide star.
This ``star'' is actually a close double, with a separation of about 0\farcs15
and an intensity ratio of 1.5:1 at $K'$ (\citealt{sto99,qui01}; but note that the image of the binary shown in an inset to Fig.~1 of Stockton et al.\ was inadvertently flipped, so North is at the bottom, East to the left). The 2$\sigma$
uncertainty in the center-of-light position of the double (or, equivalently, of either
component) is about $\pm0\farcs5$ in
right ascension and about $\pm0\farcs7$ in declination \citep{sto99}. In this paper, we describe
both new AO imaging of 3C\,294 and
new {\it Hubble Space Telescope} ({\it HST}) Fine-Guidance Sensor (FGS) observations that, together with positions from the
new UCAC2 catalog \citep{zac00, zac03}, allow us to obtain a precise
position for both components of the AO guide star and thus align the radio and
optical/infrared frames near 3C\,294 to much greater precision than before.

\section{AO Imaging}

We obtained new imaging of 3C\,294 in the $K'$ band with the Keck II AO system and 
the SCAM (slit-viewing
camera) of NIRSPEC \citep{mcL98} on 2001 May 18 (UT).  With a pixel scale of 17 mas, the $256\times256$
HgCdTe array had a total field of 4\farcs3 square. The seeing was generally good, but
with variable periods, so we selected for coadding only those images for which the eastern stellar component of
3C\,294 had a clear diffraction-limited core.  The total exposure for the
coadded image was 9000 s.

We also obtained new imaging with IRCS and the AO system on the Subaru telescope on 2003 February 17 (UT), with a total exposure of 2400 s. The main purpose of this imaging was to give an independent check to the offset of various components of 3C\,294 from the AO guide star.  Accordingly, the guide star was kept within the detector field, and the exposures were adjusted to avoid serious saturation of the two components of the star (while the peak pixels of both components did saturate slightly in most exposures, accurate centroids could still be found to a fraction of a pixel).  We also obtained imaging of the close pair of Hipparcos stars HIP63605 and HIP63607 immediately before the 3C\,294 observations in order to determine an accurate scale and orientation for the detector.  These stars have a separation of 11\farcs431.  The scale was found to be $0\farcs022516\pm0\farcs00002$ per pixel, and the detector had a rotation angle of $0\fdg165\pm0\fdg025$ with respect to north.

\begin{figure}[p]
\epsscale{1.0}
\plotone{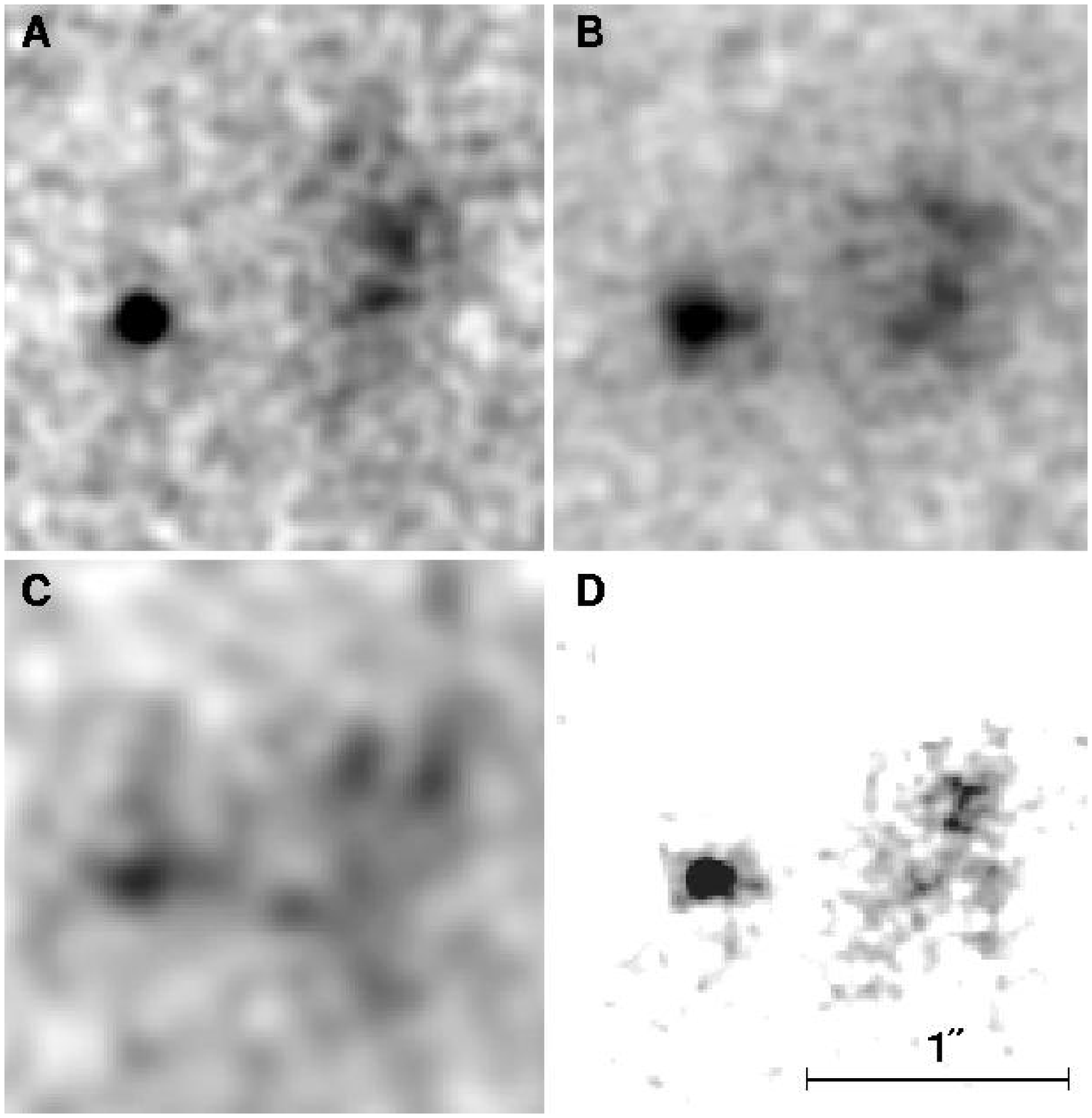}
\figcaption{Adaptive-optics imaging of 3C\,294. Panel $A$ shows the new Keck II NIRSPEC AO imaging in the $K'$ passband, with an original scale of 0\farcs01647 pixel$^{-1}$, panel $B$ shows the new Subaru IRCS AO imaging ($K'$, 0\farcs022516 pixel$^{-1}$), and panels $C$ and $D$ show previous AO imaging by \citet{sto99} ($K'$, 0\farcs03537 pixel$^{-1}$) and \citet{qui01} ($H$, 0\farcs0168 pixel$^{-1}$), respectively. Each of the images in panels $A$, $B$, and $C$ have been smoothed with a Gaussian with $\sigma=1$ pixel, and all images are reproduced at the same scale. North is up and East to the left for this and subsequent figures. See also \citet{ste02}, Fig.~4.\label{aoimagefig}}
\end{figure}

For both of these sets of observations, we used our standard iterative IR-imaging reduction procedure (see, \eg\ \citealt{sto98}). 
Figure \ref{aoimagefig} shows our new Keck and Subaru AO images, together with
images from \citet{sto99} and \citet{qui01}.

\section{{\em HST} Fine-Guidance Sensor Observations}\label{fgsobs}

\begin{figure}[p]
\epsscale{0.8}
\plotone{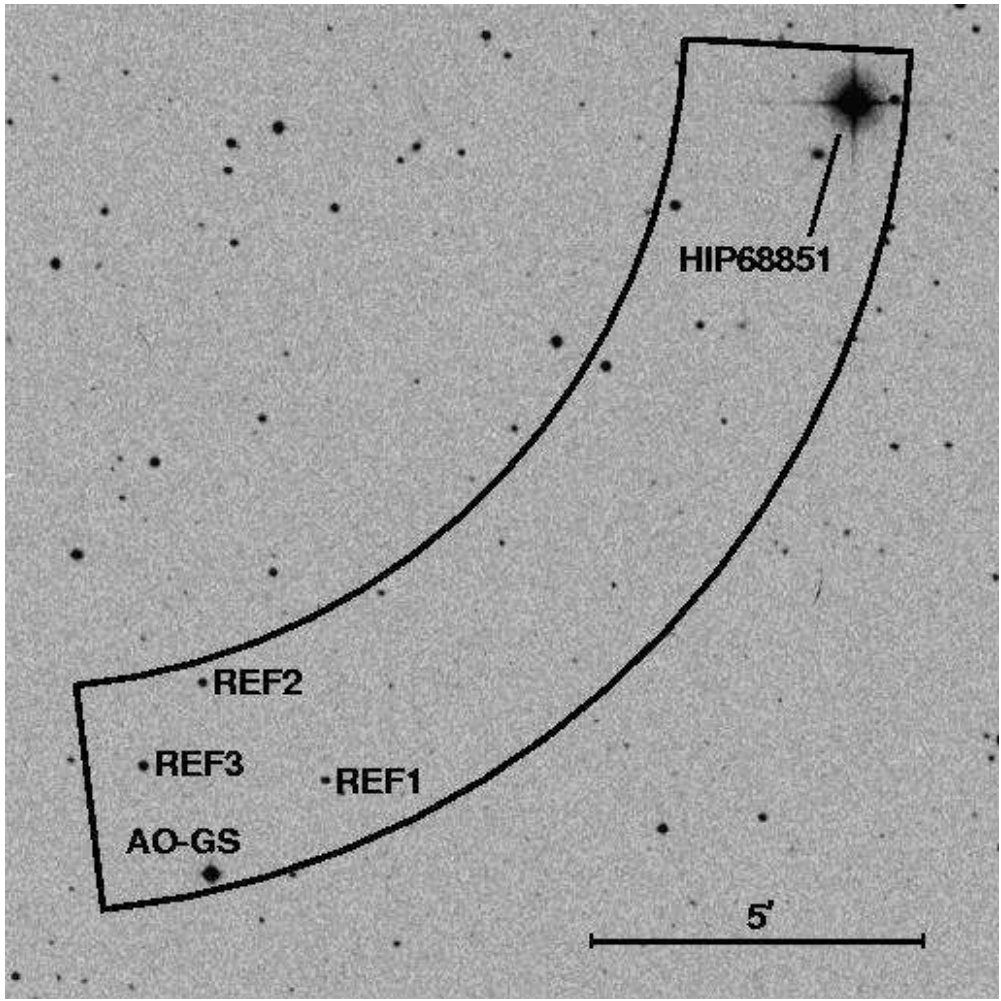}
\figcaption{Arrangement of stars observed with FGS1r. The area accessible to the FGS is shown by the enclosed solid line. The image is from the red 1st-generation Digital Sky Survey. North is up, east to the left.\label{fgsfig}}
\end{figure}

The  {\it HST} Fine-Guidance Sensor (FGS) observations were undertaken to tie the position of
the AO guide star (AO-GS) to the International Coordinate Reference System (ICRS), via the
Hipparcos star HIP68851, which could be placed within the same FGS ``pickle.''
Figure \ref{fgsfig} shows the distribution of stars across the FGS1r field of view. The FGS observations were carried out during a single orbit on 2000 December 11 UT. The FGS was used in position mode to determine the centers of HIP68851, AO-GS, and the three reference stars REF1, REF2, and REF3, as well as in transfer mode to measure the separation, position angle, and relative brightnesses of the two components of AO-GS. These measurements establish the radial separations and position angles of the stars relative to HIP68851 {\it in the FGS frame} to an accuracy of $\sim1$ mas and 0.00006 degrees respectively. Translating these quantities from the FGS frame into celestial coordinates depends critically on our knowledge of the {\it HST} roll angle, which could differ from the 
commanded roll angle by up to $\sim0\fdg04$. Over the $\sim15\arcmin$ lever arm to the Hipparcos star, this uncertainty could result in coordinate errors as large 0\farcs6 in the direction perpendicular to the line from the star in question to HIP68851. We obtain a better estimate of the actual roll angle by comparing the positions found by the FGS observations for the AO-GS, REF2, and REF3 with positions for the same stars from the U.S. Naval Observatory UCAC2 catalog \citep{zac00, zac03}. REF1 turned out to be too faint for the UCAC2 catalog; it therefore could not be used for the roll determination.

The binary components of AO-GS were well resolved by the FGS in transfer-scan mode. The binary has a separation of $0\farcs147\pm0.001$, a position angle $233\fdg25\pm0\fdg38$, and component brightness difference $\delta_V = 1.41$. 

We have assumed that the UCAC2 coordinates for the AO-GS binary are for the system's center of light. To allow a direct comparison with the FGS observations, the location of the binary's center of light in the FGS frame was determined using the binary separation, position angle, and component brightness, as measured in the FGS transfer scan observation, to determine the offset from the primary, whose position was measured by the position mode observations. The UCAC filter half-transmission points were at 5790 and 6420 \AA, while those for the FGS were at about 4350 and 7100 \AA, so the central wavelengths differed by $\sim400$ \AA. We estimate the uncertainty introduced by this difference in bandpass to be no more than 0\farcs01--0\farcs02. It is worth
emphasizing, however, that this uncertainty does not affect our final astrometry of AO-GS at
this level: it enters in quadrature along with the uncertainties in the positions of our two 
reference stars in the determination of the HST roll angle correction.

Table \ref{fgstab} compares the FGS and UCAC2 determination of the radial separation and position angle between REF2, REF3, and AO-GS and the Hipparcos star. Typical UCAC positions are accurate to 30 to 70 mas, depending upon the object's brightness and the number of times it was observed. Therefore, we can use the UCAC2 coordinates to determine the various position angles, and when averaged, to determine the correction to the commanded HST roll angle. Finally, with the HST roll angle known, the corrected position angle of each star with respect to the Hipparcos star could be determined from the FGS observations. These position angles, together with the measured radial separations, allowed the right ascension and declination of each star to be determined in the ICRS, which is anchored to the Hipparcos star.

The Hipparcos star has a significant proper motion ($\mu_\alpha, \mu_\delta$) = (+15.1, $-69.6$) mas yr$^{-1}$. This proper motion has been accounted for in determining its position at the epoch of our observations.

From the PA differences in the last column of Table \ref{fgstab}, we determine the error in the HST commanded roll angle to be $+0\fdg0260 \pm 0\fdg0005$.
\begin{center}
\begin{deluxetable}{lcccccc}
\tablewidth{0pt}
\tablecaption{Comparison of FGS and UCAC2 Positions for Stars near 3C\,294 \label{fgstab}}
\tablehead{\colhead{} & \multicolumn{2}{c}{FGS} & \multicolumn{2}{c}{UCAC2}
& \multicolumn{2}{c}{Difference} \\
\colhead{Stars} & \colhead{Separation} & \colhead{PA}
& \colhead{Separation} & \colhead{PA} & \colhead{$\delta_{\rm sep}$} & 
\colhead{$\delta_{\rm PA}$} }
\startdata
HIP--AO-GS\tablenotemark{a} & 908\farcs3226 & 139\fdg8086 & 908\farcs3219 & 139\fdg7835 & 0\farcs0007 & 0\fdg0251\\
HIP--REF2 & 788\farcs3598 & 131\fdg3455 & 788\farcs2793 & 131\fdg3185 & 0\farcs0805 & 0\fdg0270\\
HIP--REF3 & 878\farcs7950 & 132\fdg7303 & 788\farcs8165 & 132\fdg7043 & 
$-0\farcs0215$\phs & 0\fdg0260\\
\enddata
\tablenotetext{a}{AO-GS refers to the binary's center of light.}
\end{deluxetable}
\end{center}

\section{Registration of the AO Images of 3C\,294 with respect to the AO Guide Star}

In order to register the AO images to the radio coordinates, we need two kinds of coordinate data:
(1) accurate coordinates of AO-GS in the ICRS reference frame, and (2) the offset between AO-GS and some well-defined feature in the AO images. In addition, we need a precise mapping of the \citet{mcC90} radio coordinates from the B1950 VLA calibrator frame, in which they were given, to the ICRS frame.

\subsection{The Coordinates of the AO Guide Star}

Using the proper-motion-adjusted coordinates of the Hipparcos star to tie the FGS frame to the ICRS, the position of the prime component of the AO binary was determined from its FGS measured separation and position angle (using the HST corrected roll angle) relative to the Hipparcos star. Because of the large ($\sim15\arcmin$) separation between HIP68851 and AO-GS, this calculation was rigorously performed using great circles on the celestial sphere rather than by approximate planar projection onto the orthogonal ($\xi$, $\epsilon$) coordinate system of the local FGS tangential plane.

This calculation gives $14^{\rm h} 06^{\rm m} 43\fs3601 \pm 0\fs0007, +34\arcdeg 11\arcmin 23\farcs389 \pm 0\farcs009$ for the coordinates of the AO primary in the ICRS. Interestingly, the largest contribution to the errors is the uncertainty in the proper motion of HIP68851. The separation and position angle between the AO primary and the Hipparcos star is 908\farcs3202 and 139\fdg7806, respectively.

\subsection{The Offset of the 3C\,294 Field from the AO Guide Star}

We have two measures of the offset of the apparent stellar eastern component of 3C\,294 in the AO images from the center of light of the binary AO guide star: (1) from our previous CFHT imaging \citep{sto99},
and (2) from the recent Subaru imaging.  These give, respectively, ($\Delta\alpha$, $\Delta\delta$)
= (9\farcs810, 1\farcs596) and (9\farcs784, 1\farcs514). The estimated random errors in both cases are about 0\farcs01 in $\Delta\alpha$ and about 0\farcs015 in $\Delta\delta$, although there is a possibility of a small systematic error in the position angle in both measurements, which could account for the difference of $\sim0\farcs08$ in the declination offset.  We take a simple mean of these two determinations and assume that $\Delta\alpha=9\farcs797\pm0\farcs02$ and $\Delta\delta=1\farcs555\pm0\farcs05$. These values are consistent with the rough offset
of 9\farcs7 east and 1\farcs6 north given by \citet{qui01} for their AO imaging and in reasonable
agreement with the offset of $9\farcs5\pm0\farcs2$ east and $1\farcs5\pm0\farcs2$ north they
found from their lower-resolution NIRC imaging. In order to use this offset with the FGS coordinates for AO-GS, we convert it from the offset from the center of light of the binary to the offset from the primary (easternmost) component of the binary. The magnitude difference of the components at $K'$ is 0.44 mag, and we use the accurate separation of 0\farcs147 and position angle of 233\arcdeg\ determined from the FGS transfer-mode observations. The offset of the stellar object from the AO-GS primary is then $9\farcs750\pm0\farcs02$ east and $1\farcs520\pm0\farcs05$ north.

Adding this offset to the position of the AO-GS primary found above gives the position of the stellar object near 3C\,294 as $14^{\rm h} 06^{\rm m} 44\fs1459 \pm 0\fs0017, +34\arcdeg 11\arcmin 24\farcs91 \pm 0\farcs05$.

\section{The Position of the Radio Nucleus in the AO Imaging Field}

The position of the unresolved flat-spectrum radio core of 3C\,294 has been given by \citet{mcC90}, from their observations at 6 cm with the Very Large Array (VLA) in the A-array configuration. Their position for the core is $14^{\rm h} 04^{\rm m} 34\fs060 \pm 0\fs005, +34\arcdeg 25\arcmin 40\farcs00 \pm 0\farcs05$ in the reference frame defined by the VLA B1950 astrometric calibrators. Conversion of these coordinates to the ICRS (J2000) reference frame was kindly carried out for us by Jim Ulvestad. The result is $14^{\rm h} 06^{\rm m} 44\fs074 \pm 0\fs005, +34\arcdeg 11\arcmin 24\farcs95 \pm 0\farcs05$. These values are very close to those given by \citet{sto99}, who used a different procedure, but they should be more accurate; uncertainties introduced by the conversion should be $<1$ mas (J. Ulvestad, private communication).

\begin{figure}[!tb]

\epsscale{1.0}
\plotone{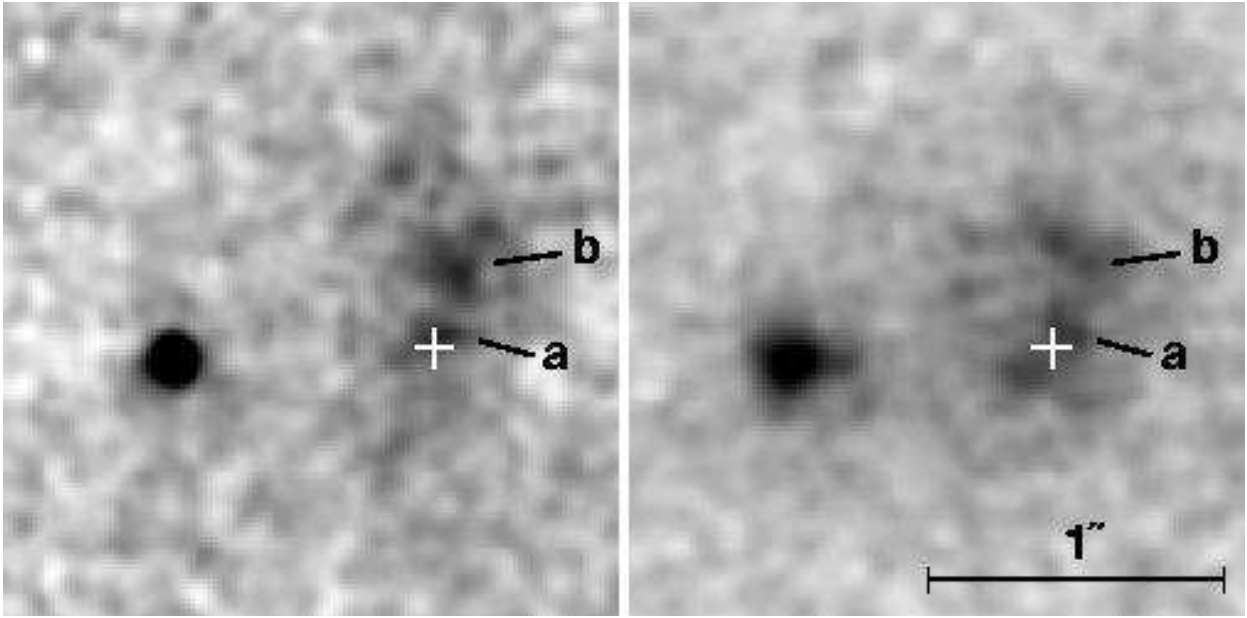}
\figcaption{Location of the radio core on AO images. The left panel shows the Keck II NIRSPEC image, and the right panel shows the Subaru IRCS image. The white crosses show the location and estimated $1\sigma$ error bars for the radio core. Two components of the galaxy that appear to be present in both images are labelled, $a$ being the likely identification of the nucleus.\label{ao_core}}
\end{figure}

With our precise mapping of the AO imaging onto the ICRS reference frame, we can now locate the position of the radio core on the AO $K'$-band image. It lies $0\farcs89\pm0\farcs07$ west and $0\farcs04\pm0\farcs07$ north of the stellar object, as shown in Fig.~\ref{ao_core}.  The radio core, which presumably marks the nucleus, lies very close to one of the brightness peaks of what \citet{qui01} have called ``the western component'' of 3C\,294. On the other hand, this position of the radio core is clearly inconsistent with its being identified with the stellar object, as was suggested by \citet{qui01}.

One of the most striking features of Fig.~\ref{aoimagefig} is the difference in appearance of the various AO images.  All show a stellar component and a diffuse component about 1\arcsec\ to the west of it, but there is little detail that convincingly reproduces from image to image. To be sure, panel $C$ is a rather noisy image, and panel $D$ \citep{qui01} is an $H$-band image (the other 3 are all $K'$), so we should probably concentrate on comparing the Keck II NIRSPEC image $A$ and the Subaru IRCS image $B$. These are reproduced in Fig.~\ref{ao_core}, where we have marked two features that seem to be present in both images. Most of the other apparent features in one or the other of the images are likely due to speckle noise.

The strongest of the ``real'' features (aside from the stellar object to the east), labeled $a$ in Fig.~\ref{ao_core}, lies within the error bars of the position of the radio core and is plausibly to be identified with the nuclear region of 3C\,294. The position of object $a$ falls near a local minimum in the $H$-band image of \citet{qui01} (see Fig.~\ref{aoimagefig}$D$; this weakness may indicate that the nucleus is strongly obscured at shorter wavelengths. Such obscuration would be consistent with the conclusion of \citet{fab03} that the central hard X-ray source in 3C\,294 is a highly obscured quasar with a luminosity of $\sim10^{45}$ erg s${-1}$.

\section{The Eastern Stellar Object}

If the radio nucleus  lies within the diffuse nebulosity, what is the nature of stellar object to the east? The density of star-like objects of similar or greater brightness at $K$ in the field of 3C\,294 is on the order of 10 per square arcmin, so it would be quite unusual for there to be an unrelated object within 1\arcsec\ of the nucleus of 3C\,294. On the other hand, there is no compelling evidence for significant nebulosity associated with the stellar object, so it {\it could} be a projected Galactic subluminous star---we are acutely aware of misinterpretations arising from stars projected on, \eg\ the radio galaxy 3C\,368 \citep{ham91,sto96} and the quasar PKS\,2135$-$147 \citep{can97}.

\begin{figure}[!tb]
\epsscale{1.0}
\plotone{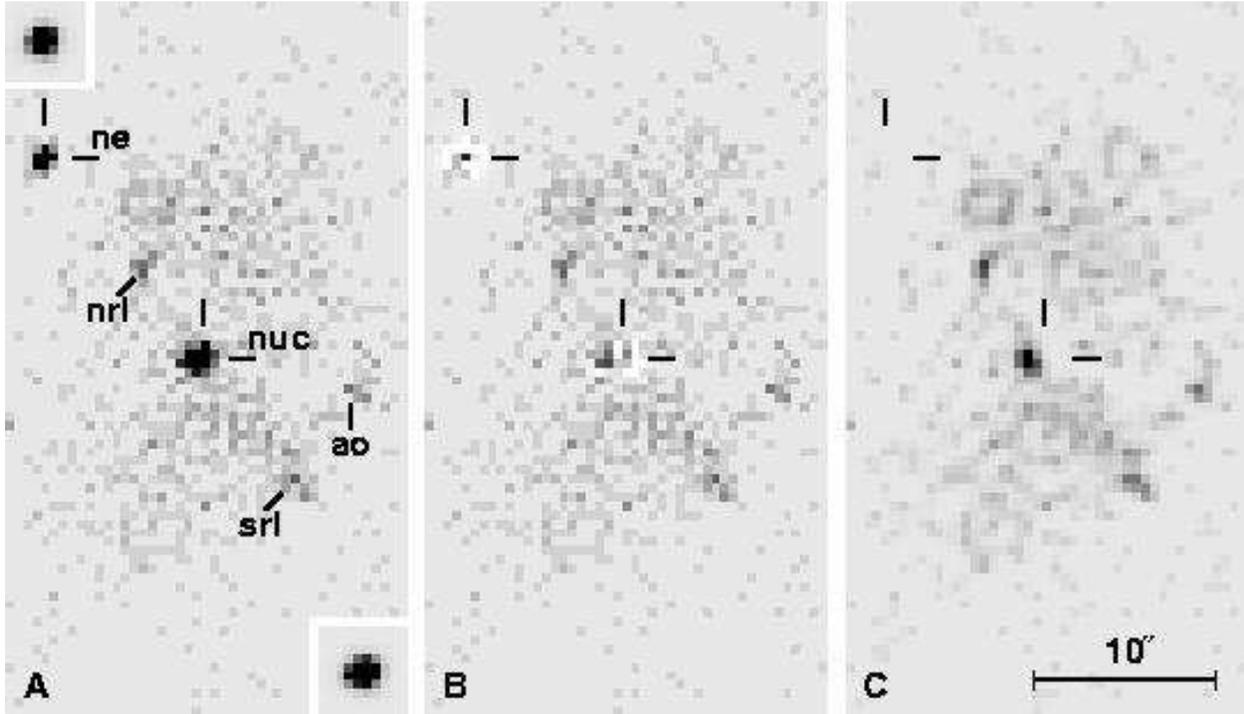}
\figcaption{A deep (192 ks) Chandra image of the region immediately around 3C\,294. Panel $A$ shows the image, filtered to include events from 0.5--6 keV. The upper-left and lower-right insets to panel $A$ show model PSFs generated by the CIAO ChaRT ray tracing procedure at the positions of the peaks of the northeastern point source (ne) and the nuclear point source (nuc). Sources corresponding with other features identified by \citet{fab03} as the northern and southern radio hotspots and the star we used for AO guiding are also indicated. Panel $B$ shows the result of subtracting the model PSFs from the original image; note the excess flux to the east of the nucleus. Panel $C$ shows the result of a two-component deconvolution using a subsampled PSF and 
removal of the nuclear and northwestern point sources, as described in the text. The short lines to 
the north and west of northeastern and nuclear sources show the positions of the peaks.\label{ximage}}
\end{figure}

In the case of 3C\,294, we have some additional evidence on the nature of the stellar object. We have obtained from the Chandra Data Archive the 192 ks integration used by \citet{fab03} on 3C\,294 to map the surrounding extended X-ray emission. The region immediately around 3C\,294 is shown in Fig.~\ref{ximage}$A$. A few discrete sources besides the nuclear component are visible, including a faint source at the position of the AO guide star and a bright point source northeast of the nucleus, which \citet{fab03} interpret as a Seyfert II nucleus with a redshift close to that of 3C\,294. The distance between the X-ray peaks corresponding to the AO guide star and the nuclear component agrees, perhaps somewhat fortuitously (given the image scale and count statistics) with that determined from the AO imaging to within 0\farcs1.

The nuclear component itself appears elongated in the east-west direction. Part of this elongation is due to aberrations because of the off-axis position of the source, as can be seen in the model PSFs shown in the insets to Fig.~\ref{ximage}$A$, but the subtraction shown in Fig.~\ref{ximage}$B$ shows that part of the elongation is actually due to a second source centered $\sim2$ pixels ($\sim1\arcsec$) to the east, \ie\ essentially coincident with the stellar object. 

We have also attempted deconvolution of the X-ray image, using a PSF kernel generated for the position and spectral-energy-distribution of the nucleus over the 0.5 to 6 keV band, using the 
CIAO\footnotemark[2] procedure ChaRT to trace the rays and MARX\footnotemark[3] to generate a pseudo event file. The PSF was generated on a grid subsampled by a factor of 4 with respect to the Chandra ACIS pixel scale. We used the procedure {\it cplucy} \citep{hoo99}, which carries out a
two-channel deconvolution. One channel includes the point sources (in our case, we include only the nuclear and the northeastern sources), which are modeled and removed. The other channel contains the ``background,'' which includes all sources not designated as point sources. This
background channel is deconvolved subject to an entropy constraint. The subsampled 
deconvolved image was resampled back to the ACIS scale. This result is shown in Fig.~\ref{ximage}$C$; it also shows a source at the position of the eastern stellar object.
\footnotetext[2]{http://cxc.harvard.edu/ciao/}
\footnotetext[3]{http://space.mit.edu/CXC/MARX/}

Thus, we have two point x-ray sources within about 1\arcsec\ of each other, strengthening the likelihood that the objects are closely related. In addition, we can use the X-ray---optical flux ratio
to constrain the nature of the eastern component. The flux ratio is defined by 
${\rm log} (f_x/f_V)={\rm log} f_x+m_V/2.5 + 5.37$, where $f_x$ is the X-ray flux in
erg cm$^{-2}$ s$^{-1}$ within the 0.3--3.5 keV range. The eastern stellar object has $K\sim21.2$ (\citealt{qui01} give $K=21.7$ but mention some uncertainty due to nonphotometric conditions), and its X-ray flux in the 0.3--3.5 keV band is $\sim10^{-15}$ erg cm$^{-2}$ s$^{-1}$. Among normal stars, the highest ratios of $f_x/f_V$ are found for M stars, for which they range from about 0.001 to about 0.3 (\eg\ \citealt{mac88}). If the object were an M star, it would have a $V\!-\!K$ color of $\gtrsim4$, so an $m_V\gtrsim25$. With the X-ray flux given above, this $m_V$ results in 
$f_x/f_V\gtrsim2.3$, well above the likely range for normal stars. While earlier-type stars have less-red colors, so their $m_V$ would be smaller, their expected values of $(f_x/f_V)$ drop even faster, so the discrepancy would be even larger.

On the other hand, if the object were an AGN at the redshift of 3C\,294, it would typically have a $V\!-\!K$ color of $\sim2$, so $m_V\sim23$ and $f_x/f_V\sim0.4$, a ratio that is within the expected range for AGN. The most likely explanation for the duplicity in the Chandra image is therefore that there are two active nuclei associated with 3C\,294: one associated with the compact radio nucleus and heavily obscured at rest-frame optical wavelengths and in soft X-rays \citep{fab03}, the other associated with the stellar object 0\farcs9 ($\sim8$ kpc) east of the radio source and intrinsically less powerful, but having relatively little extinction. If we assume no extinction for the eastern nucleus and assume the estimate of \citet{fab03} for the intrinsic X-ray flux of the radio nucleus, the eastern nucleus has an intrinsic X-ray luminosity roughly an order of magnitude lower 
than that of the radio nucleus. There is no indication of radio emission from this second active nucleus in the 6 cm VLA map of \citet{mcC90}, but this fact alone cannot set very stringent limits on its radio/X-ray flux ratio.

\section{Summary and Discussion}

The position of the radio nucleus of 3C\,294 determined here places it $\sim0\farcs8$ north of the position given by \citet{sto99} and $\sim0\farcs9$ west of the position given by \citet{qui01}. Within the errors of the determination, the nucleus is coincident with a modest peak within the diffuse component seen in the $K$-band imaging data. While this position of the radio nucleus undercuts the specific argument made by \citet{qui01} that 3C\,294 is a merger in progress, since the radio source can no longer be identified with the eastern stellar component, their conclusion is nevertheless reaffirmed by the fact that both the radio nucleus and the eastern stellar object appear to be X-ray sources.

The apparent presence of two active nuclei in such close proximity places 3C\,294 within a
a small, but important, class. A recent cottage industry has developed in mining surveys for
gravitationally lensed QSOs to extract double QSOs that are {\it not} the result of gravitational 
lensing,
but, instead, are true binaries \citep[\eg][]{mor99}. Such cases are important for at least two
reasons: (1) If they are found as subsets of large survey whose selection properties are well 
determined, such as the Large Bright Quasar Survey (LBQS) or the Sloan Digital Sky Survey 
(SDSS), their statistics can provide evidence for the importance of interactions and mergers
in triggering nuclear activity \citep{djo91,koc99}; and (2) with a sufficiently large sample, they 
can provide evidence on mean lifetimes of QSO activity \citep{mor99}.
3C\,294, with active nuclei at a projected distance $\lesssim8$ kpc, has one of the smallest 
projected separations yet found. Other examples with good credentials as true close binaries 
include LBQS\,0103$-$2753 \citep[$z=0.85$, $d=0\farcs3=2.3$ kpc;][]{jun01}, 
FIRST J164311.3+315618 \citep[$z=0.59$, $d=2\farcs3=15$ kpc;][]{bro99}, 
SDSS\,J233646.2$-$010732.6 
\citep[$z=1.285$, $d=1\farcs67=15$ kpc;][]{gre02}, and LBQS\,0015+0239
\citep[$z=2.45$, $d=2\farcs2=18$ kpc;][]{imp02}. \citet{sma03} have noted that cases of submillimeter-bright galaxies that are detected as pairs of X-ray sources at separations of
a few tens of kpc are much more common than would be expected by chance, and they have 
suggested that these represent early stages of mergers of two galaxies, each of which hosts a 
supermassive black hole. They also point out that \citet{rav02} have shown that the flattened 
stellar distribution seen in the
cores of many cluster ellipticals can be produced by the dynamical action of a close
supermassive-black-hole binary. Very close pairs of active nuclei, such as 3C\,294, likely give us
snapshots of stages in this process between that seen in the X-ray double submillimeter sources
and the final formation of a tight binary black hole, by which time one or both nuclei may have
ceased to show activity. Together with the overdensity of red galaxies
around it found by \citet{tof03}, this result gives us considerable confidence that in 3C\,294 we are 
indeed witnessing a stage in the formation of a dominant cluster galaxy.

\acknowledgments

We are very grateful to Norbert Zacharias
of the U. S. Naval Observatory for providing coordinates from the UCAC2 survey
in advance of publication. We are also indebted to Christ Ftaclas, Pat Henry, Mike Liu, and Fred Lo for discussions, to Andi Mahdavi and Andrea Prestwich for advice on using CIAO,
and to
Jim Ulvestad for the conversion from the VLA B1950 calibrator frame to the ICRS
J2000 frame. Support for this work was provided by NASA through Grant No.\ GO-08315.01-A
from the Space Telescope Science Institute, which is operated by AURA, Inc.,
under NASA Contract No.\ NAS 5-26555. G.~C.\ was supported in part under the auspices 
of the U.S.\ Department of Energy, National Nuclear Security 
Administration by the University of California, Lawrence Livermore 
National Laboratory under contract No.\ W-7405-Eng-48. S.~E.~R.\ was supported
in part by NASA LTSA grant NAG 5-10762.
This research has made use of the NASA/IPAC Extragalactic Database (NED) 
which is operated by the Jet Propulsion Laboratory, California Institute of 
Technology, under contract with the National Aeronautics and Space 
Administration.

\clearpage

\end{document}